\documentclass[floatfix,aps,prl,twocolumn, superscriptaddress]{revtex4-2}
\usepackage{upgreek}
\usepackage{graphicx}
\usepackage[]{epsfig}
\usepackage{times,amsmath,amssymb}

\usepackage{color}
\usepackage[colorlinks = true,
            linkcolor = blue,
            urlcolor  = blue,
            citecolor = blue,
            anchorcolor = blue]{hyperref}

\begin{document}

\title{Quantum dot transistors based on CVD-grown graphene nano islands}

\author{Takumi Seo}
\email[These authors contributed equally to this work]{}
\affiliation{Research Institute of Electrical Communication, Tohoku University, 2-1-1 Katahira, Aoba-ku, Sendai 980-8577, Japan}
\affiliation{Department of Electronic Engineering, Graduate School of Engineering, Tohoku University, Aoba 6-6-05, Aramaki, Aoba-Ku, Sendai 980-8579, Japan}

\author{Motoya Shinozaki}
\email[These authors contributed equally to this work]{}
\affiliation{WPI Advanced Institute for Materials Research, Tohoku University, 2-1-1 Katahira, Aoba-ku, Sendai 980-8577, Japan}
\affiliation{Research Center for Materials Nanoarchitechtonics (MANA), National Institute for Material Science (NIMS),
1-2-1 Sengen, Tsukuba 305-0047, Japan}

\author{Akiko Tada}
\email[These authors contributed equally to this work]{}
\affiliation{WPI Advanced Institute for Materials Research, Tohoku University, 2-1-1 Katahira, Aoba-ku, Sendai 980-8577, Japan}

\author{Yuta Kera}
\affiliation{Research Institute of Electrical Communication, Tohoku University, 2-1-1 Katahira, Aoba-ku, Sendai 980-8577, Japan}
\affiliation{Department of Electronic Engineering, Graduate School of Engineering, Tohoku University, Aoba 6-6-05, Aramaki, Aoba-Ku, Sendai 980-8579, Japan}

\author{Shunsuke Yashima}
\affiliation{Research Institute of Electrical Communication, Tohoku University, 2-1-1 Katahira, Aoba-ku, Sendai 980-8577, Japan}
\affiliation{Department of Electronic Engineering, Graduate School of Engineering, Tohoku University, Aoba 6-6-05, Aramaki, Aoba-Ku, Sendai 980-8579, Japan}

\author{Kosuke Noro}
\affiliation{Research Institute of Electrical Communication, Tohoku University, 2-1-1 Katahira, Aoba-ku, Sendai 980-8577, Japan}
\affiliation{Department of Electronic Engineering, Graduate School of Engineering, Tohoku University, Aoba 6-6-05, Aramaki, Aoba-Ku, Sendai 980-8579, Japan}
\affiliation{Research Center for Materials Nanoarchitechtonics (MANA), National Institute for Material Science (NIMS),
1-2-1 Sengen, Tsukuba 305-0047, Japan}

\author{Takeshi Kumasaka}
\affiliation{WPI Advanced Institute for Materials Research, Tohoku University, 2-1-1 Katahira, Aoba-ku, Sendai 980-8577, Japan}

\author{Azusa Utsumi}
\affiliation{WPI Advanced Institute for Materials Research, Tohoku University, 2-1-1 Katahira, Aoba-ku, Sendai 980-8577, Japan}

\author{Takashi Matsumoto}
\affiliation{S-Technology Development Center, Tokyo Electron Technology Solutions Limited, 650 Mitsuzawa, Hosaka-cho, Nirasaki, Yamanashi 407-0192, Japan}

\author{Yoshiyuki Kobayashi}
\affiliation{S-Technology Development Center, Tokyo Electron Technology Solutions Limited, 650 Mitsuzawa, Hosaka-cho, Nirasaki, Yamanashi 407-0192, Japan}

\author{Tomohiro Otsuka}
\email[]{tomohiro.otsuka@tohoku.ac.jp}
\affiliation{WPI Advanced Institute for Materials Research, Tohoku University, 2-1-1 Katahira, Aoba-ku, Sendai 980-8577, Japan}
\affiliation{Research Institute of Electrical Communication, Tohoku University, 2-1-1 Katahira, Aoba-ku, Sendai 980-8577, Japan}
\affiliation{Department of Electronic Engineering, Graduate School of Engineering, Tohoku University, Aoba 6-6-05, Aramaki, Aoba-Ku, Sendai 980-8579, Japan}
\affiliation{Research Center for Materials Nanoarchitechtonics (MANA), National Institute for Material Science (NIMS),
1-2-1 Sengen, Tsukuba 305-0047, Japan}
\affiliation{Center for Science and Innovation in Spintronics, Tohoku University, 2-1-1 Katahira, Aoba-ku, Sendai 980-8577, Japan}
\affiliation{Center for Emergent Matter Science, RIKEN, 2-1 Hirosawa, Wako, Saitama 351-0198, Japan}


\begin{abstract}
Graphene nanoislands (GNIs) are one of the promising building blocks for quantum devices owing to their unique potential.
However, direct electrical measurements of GNIs have been challenging due to the requirement of metal catalysts in typical synthesis methods.
In this study, we demonstrate electrical transport measurements of GNIs by using microwave plasma chemical vapor deposition, which is a catalyst-free method to deposit graphene directly on SiO$_2$ substrates.
This approach enables the fabrication of metal electrodes on GNIs, allowing us to measure their quantum transport properties.
At low temperatures, one of our devices shows clear Coulomb diamonds with twofold degeneracy, indicating the formation of quantum dots and the vanishing of valley degeneracy.
The charge state of the GNI is also modulated by a local side gate, and the tunneling coupling between leads and quantum dots is modulated by changing contact area and metal materials.
These results provide device design guidelines toward GNI-based quantum devices for next-generation computing.
\end{abstract}

\maketitle

Graphene, a two-dimensional carbon material, has attracted attention owing to its remarkable electronic properties~\cite{novoselov2004electric}, including high carrier mobility, low spin-orbit coupling~\cite{Huertas2006spin}, and so on. 
These unique characteristics make graphene a promising candidate for high-performance quantum devices~\cite{trauzettel2007spin}, such as quantum dots with long relaxation time~\cite{banszerus2022spin, garreis2024long}. 

Quantum dots, called artificial atoms owing to their discrete energy levels and controllable electron numbers, have been investigated from both viewpoints of fundamental physics and applicationss~\cite{tarucha1996shell, kouwenhoven1997excitation, kouwenhoven2001few}.
Although optical measurements have been the primary tool to investigate physically defined nanoscale quantum dots such as colloidal systems~\cite{press2008complete, warburton2013single, tian2023review}, advances in nanotechnology have enabled the fabrication of electrical contacts to individual quantum dots, allowing for direct transport measurements.
This electrical approach to colloidal quantum dot systems has demonstrated their potential as single-electron transistors operating at room temperatures, opening new frontiers in quantum device applications~\cite{shibata2023single}.

To create such nanoscale structures in graphene, chemical vapor deposition (CVD), typically known as a large-scale synthesis method, can be utilized~\cite{yu2008graphene, reina2009large, li2009large, ago2012catalytic}.
The CVD growth is utilized not only to fabricate graphene nanoribbon~\cite{kato2012site, suzuki2016wafer, kato2022scalable} but also self-assembled graphene nanoislands (GNIs) which can be unexpectedly formed with nanoscale dimensions.
The electronic properties of these GNIs have been investigated by scanning tunneling microscopy (STM), revealing discrete density of states~\cite{phark2012scanning} and pseudo-magnetic field effect~\cite{levy2010strain, settnes2016}. 
However, while these STM studies suggest the potential of graphene nanoislands for quantum devices, the actual device fabrication and measurement of the devices have been challenging.
This is because typical CVD methods to deposit graphene require metal substrate catalysts, which induce parallel current paths through the substrates.
Although catalyst-free CVD methods have been previously explored and electrical measurements of larger single-layer graphene domains have been achieved~\cite{wei2013critical}, quantum effects in these structures have not been reported.

In this study, we demonstrate direct electrical measurements of graphene nanoislands deposited by a catalyst-free CVD growth method directly on Si/ SiO$_2$ substrates. 
This approach enables us to fabricate metal electrodes on substrates contacting fine structures of graphene, allowing us to measure quantum transport properties of GNI at low temperatures.

\begin{figure*}
\begin{center}
  \includegraphics{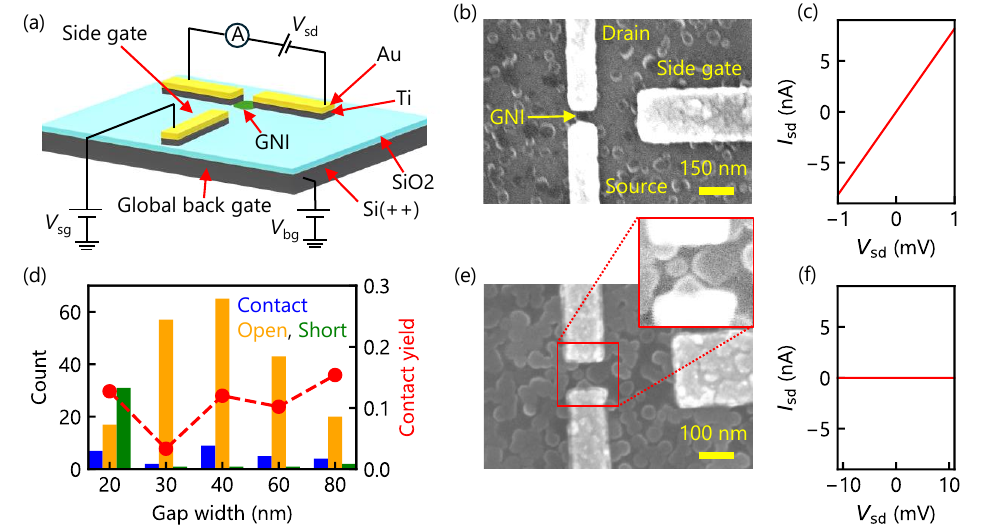}
  \caption{(a) Schematic of the device structure.
  (b) Scanning electron microscope (SEM) image of a fabricated device where both electrodes are in contact with the GNI.
  (c) Typical $I_\mathrm{sd}$-$V_\mathrm{sd}$ characteristics for devices classified as  contacts.
  (d) Number of devices showing electrical contact, open, and short to GNIs.
  (e) SEM image of a device showing an electrical open circuit.
  (f) Typical $I_\mathrm{sd}$-$V_\mathrm{sd}$ characteristics for an electrical open.}
  \label{fig1}
\end{center}
\end{figure*}

Figure ~\ref{fig1}(a) illustrates the concept of electrical measurements. 
We perform microwave plasma CVD to synthesize graphene layers on 300~mm Si/SiO$_2$ substrates below the temperature of 400$^\circ$C~\cite{yamada2013low}.
The Ti/Au source and drain electrodes are deposited on GNIs directly grown on Si/SiO$_2$ substrates by and electron beam deposition and lithography.
The width of gaps between source and drain electrodes is designed to range from 20 to 80~nm. We also align a side gate electrode near the gap.
GNIs are randomly created on the substrates and contacted in the electrode gaps, as shown in the scanning electron microscope (SEM) image of a typical device in Fig.~\ref{fig1}(b). 
In successful cases, we can obtain a source-drain current $I_\mathrm{sd}$ at room temperature as shown in Fig.~\ref{fig1}(c).
Figure~\ref{fig1}(d) summarizes the number of devices achieving electrical contact to GNIs. 
The device states are classified based on room temperature measurements without gate voltage. 
Devices with resistance above 100~M$\Omega$ are considered electrically open, those with resistance below tens of k$\Omega$ are classified as electrical shorts, and contacts typically show resistance ranging from 0.1 to several M$\Omega$.
The contact yield is approximately 0.1 for all gap sizes. 
In particular, gaps designed as 20~nm show 31 cases of electrical shorts due to their narrow width. 
Most devices result in electrical open circuits, with typical SEM images and transport characteristics shown in Figs.~\ref{fig1}(e) and (f). 
As clearly seen, GNIs are not connected to each other, resulting in electrical open.
These results highlight a fundamental challenge in the electrical measurement of GNIs. 
We need smaller GNIs to observe quantum effects, achieving electrical contact to such small GNIs requires narrower electrode gaps, which increases the probability of electrical shorts as demonstrated in this study. 
For our devices, we fabricate the electrodes using electron beam lithography. 
Here, alternative approaches such as electrical break junction techniques could potentially provide more precise control over nanogap formation~\cite{park1999fabrication, strachan2005controlled, dubois2018massively}, offering a promising direction for future device fabrication.

\begin{figure}
\begin{center}
  \includegraphics{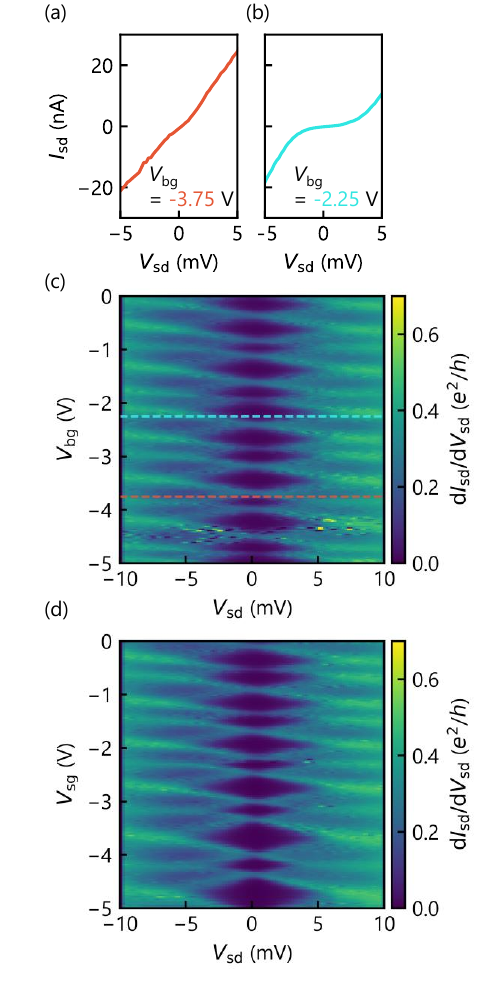}
  \caption{(a) $I_\mathrm{sd}$-$V_\mathrm{sd}$ characteristics at $V_\mathrm{bg}=-3.75$~V and (b) at $V_\mathrm{bg}=-2.25$~V measured at 2~K.
  (c) Differential conductance $\mathrm{d}I_\mathrm{sd}/\mathrm{d}V_\mathrm{sd}$ as a function of $V_\mathrm{sd}$ and $V_\mathrm{bg}$, and (d) with $V_\mathrm{sd}$ and $V_\mathrm{sg}$.}
  \label{fig2}
\end{center}
\end{figure}

To explore quantum effects, we measure the transport characteristics at a low temperature of 2~K using a helium decompression refrigerator. 
Figures~\ref{fig2}(a) and (b) show $I_\mathrm{sd}$-$V_\mathrm{sd}$ curves at back gate voltages $V_\mathrm{bg}$ of -3.75 and -2.25~V, respectively. 
While we observe an Ohmic-like behavior at $V_\mathrm{bg}=-3.75$~V, a suppressed current region appears at $V_\mathrm{bg}=-2.25$~V.
Figure~\ref{fig2}(c) shows $\mathrm{d}I_\mathrm{sd}/\mathrm{d}V_\mathrm{sd}$ normalized by the conductance quantum $e^2/h$ as a function of $V_\mathrm{sd}$ and $V_\mathrm{bg}$. 
A clear diamond shape appears in the figure, known as a Coulomb diamond, indicating quantum dot formation.

We also measure the Coulomb diamond by sweeping the side gate voltage $V_\mathrm{sg}$, as shown in Fig.~\ref{fig2}(d), and observe similar results to those obtained by sweeping $V_\mathrm{bg}$. 
Side gate modulation provides flexibility in device design, allowing for back-gate-free structures using undoped Si substrates~\cite{johmen2023radio, Reuckriegel_electric2024}.
Such structures allow us to perform radio-frequency measurements, which are important techniques for investigating quantum dynamics~\cite{qin2006radio, reilly2007fast, barthel2009rapid}. 
Our results suggest that GNI quantum dots can achieve quantum state control and readout by utilizing high-speed electronic techniques.

\begin{figure}
\begin{center}
  \includegraphics{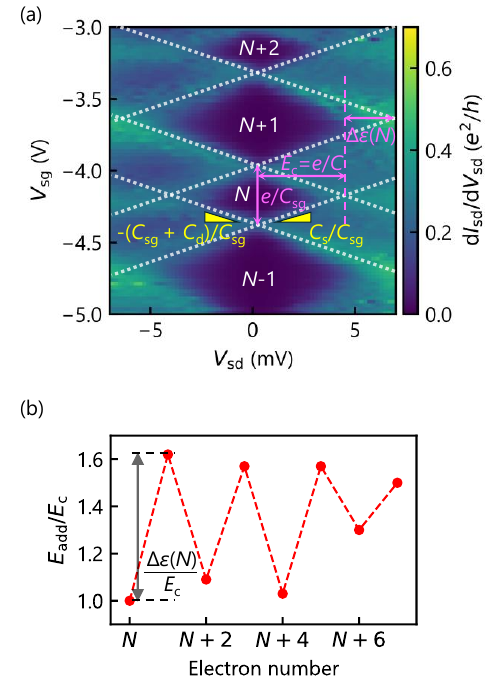}
  \caption{(a) Enlarged view of Coulomb diamonds obtained by sweeping $V_\mathrm{sg}$, and analysis guideline for the constant-interaction model.
  (b) The addition energy normalized by the charging energy as a function of electron number from $N$ to $N+7$.}
  \label{fig3}
\end{center}
\end{figure}

In the case of this device, the size of the Coulomb diamonds changes periodically in both measurements when sweeping $V_\mathrm{bg}$ and $V_\mathrm{sg}$.
We analyze the smallest Coulomb diamond shown in Fig.~\ref{fig2}(c) using the constant-interaction model~\cite{kouwenhoven2001few}, as illustrated in Fig.~\ref{fig3}(a).
Here, $C_\mathrm{g}$ is the capacitance between the GNI quantum dot and gate electrode, $C_\mathrm{s}$ the capacitance to the source electrode, and $C_\mathrm{d}$ the capacitance to the drain electrode, and $C$ the total capacitance of the device.
Full depletion of carriers in GNIs cannot be observed even when applying higher gate voltages, which makes it difficult to count the exact number of carriers.
Therefore, we denote $N$ as the number of carriers, assuming electron transport.
The charging energy $E_\mathrm{c}$ is obtained from the width of the diamond along the $V_\mathrm{sd}$ axis, giving $E_\mathrm{c} = e^2/C$, where $e$ is the elementary charge.
The lever arm $\alpha$ is calculated from the ratio of $C_\mathrm{g}$ and $C$, which converts the gate voltage to the energy scales.
The extracted values are summarized in Table.~\ref{table1}.
The values of $C_\mathrm{sg}$ show similar capacitance to $C_\mathrm{bg}$ and can be controlled by the side gate electrode design. 
The similar values of $C_\mathrm{s}$ and $C_\mathrm{d}$ indicate that both electrodes are deposited symmetrically on the GNI, which is reflected in the symmetric shape of the Coulomb diamonds.
From $C_\mathrm{bg}$, we estimate the quantum dot size $D$ to be 73~nm by assuming a parallel circular plate model following $C_\mathrm{bg}=\varepsilon_0\varepsilon_\mathrm{r}\pi D^2/4d$, where $\varepsilon_0$ is the permittivity of vacuum, $\varepsilon_\mathrm{r}=3.9$ and $d=285$~nm are the dielectric constant and the thickness of SiO$_2$, respectively.
This value is consistent with the typical size of GNIs observed in our substrates.

The difference between the shape of consecutive diamonds provides the orbital energy spacing $\Delta \varepsilon (N)$, which can be extracted when the diamond size alternates between larger and smaller values.
In graphene quantum dots, both valley and spin degrees of freedom typically result in a fourfold periodic pattern in the Coulomb diamond size~\cite{Kurzmann2021}.
The larger diamonds appear when electrons begin to occupy the next orbital level, requiring additional orbital energy on top of the charging energy.
Figure~\ref{fig3}(b) shows the addition energy $E_\mathrm{add}$ normalized by $E_\mathrm{c}$ for each electron number, where $E_\mathrm{add}$ is evaluated from the width of each diamond.
We observe that $E_\mathrm{add}$ alternates with the addition of every two electrons, indicating twofold degeneracy.
This suggests that the valley degeneracy has vanished in our GNI quantum dots, likely due to the low crystallinity of the GNIs.
The size of the quantum dot is also estimated by considering a harmonic oscillator potential in the GNI.
In this potential, the wave function is assumed to be Gaussian, and its spatial dispersion $2\hbar/\sqrt{m^*\Delta \varepsilon(N)}$ corresponds to the dot size, where $m^*=0.022m_0$ is the effective mass of bilayer graphene~\cite{Li2016effective}.
From this model, we evaluate the typical dot size to be 71~nm, which is in good agreement with the value obtained from the parallel circular plate model.

\begin{table}
  \caption{Summary of electrostatic capacitances and charging energy. The units of each capacitance and energy are given as aF and meV, respectively.}
  \label{table1}
  \vspace{2mm}
  \centering
  \begin{tabular}{ccccccc}
    & $C_\mathrm{g}$  & $C_\mathrm{s}$  & $C_\mathrm{d}$  & $C$ & $\alpha = C_\mathrm{g}/C$ & $E_\mathrm{c}$\\
  \hline\hline
  Back gate &  0.5&  17&  14&  35&  0.014& 4.6\\
  \hline
  Side gate &  0.4&  18&  19&  38& 0.011&  4.3\\
  \hline
  \end{tabular}
\end{table}

\begin{figure}
\begin{center}
  \includegraphics{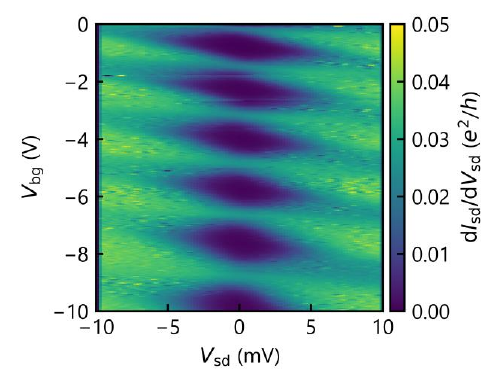}
  \caption{Coulomb diamonds obtained from another device with Cr/Au contacts by sweeping $V_\mathrm{bg}$.}
  \label{fig4}
\end{center}
\end{figure}

Figure~\ref{fig4}(a) shows the Coulomb diamonds of another device obtained by sweeping $V_\mathrm{bg}$.
We employ Cr/Au electrodes for this device to investigate the contact metal dependence of transport characteristics.
Clear diamonds are observed similar to the previously presented results, providing an estimated dot size of approximately 33~nm assuming the parallel circular plate model.
All diamonds show almost identical shapes despite the dot size being small enough to expect cyclic changes in diamond size reflecting orbital energy spacing.
The origin of this behavior remains unclear and requires further investigation.
The shape of the diamonds is slightly rounded compared to the device shown in Fig.~\ref{fig2}, indicating an increased tunneling rate between the quantum dot and lead electrodes.
This change might be caused by the difference in work function between Cr and Ti, which modulates the barrier between the dot and electrodes.
However, the device conductance is smaller than that of the previous device. 
The possible reason may be the asymmetry in tunnel coupling.

In conclusion, we have demonstrated electrical transport measurements of GNIs directly grown on SiO$_2$ substrates by microwave plasma CVD method.
Owing to the catalyst-free CVD growth, we fabricate metal electrodes directly on the GNIs.
Low-temperature measurements reveal clear Coulomb diamonds, indicating the formation of quantum dots.
The Coulomb diamonds of the presented device show twofold degeneracy instead of the fourfold degeneracy typically observed in graphene quantum dots, suggesting the vanishing of valley degeneracy due to the low crystallinity of GNIs.
The dependence on contact metals shows that the tunneling coupling between leads and quantum dots can be controlled through material selection.
Our results show that GNI quantum dots can utilize radio-frequency techniques and possess tunability of tunnel coupling by optimizing device structure design.
Furthermore, all processes to fabricate devices, including graphene synthesis, are performed below 400$^\circ$C, making our approach suitable for semiconductor integration.
These findings provide not only fundamental insights into the quantum transport properties of GNIs but also device design guidelines to realize quantum computing.

\section{Acknowledgements}

We thank RIEC Fundamental Technology Center and the Laboratory for Nanoelectronics and Spintronics for technical support. 
Part of this work is supported by 
Grants-in-Aid for Scientific Research (21K18592, 23K26482, 23H04490, 25H01504, 25H02106), 
CREST (JPMJCR23A2), JST, 
and FRiD Tohoku University.

\section{Competing interests}
The authors declare no competing interests.

\bibliography{reference.bib}

\end{document}